\documentstyle[preprint,prd,aps,eqsecnum]{revtex}
\tighten
\begin{document}
\draft
\preprint{gr-qc/9409002\qquad UCSBTH-94-32}
\begin{title}
	{\sf Qualitative Analysis of Brans-Dicke Universes with a
			Cosmological Constant}
\end{title}
\author			{\sc Shawn J.~Kolitch
	\footnote{E-Mail Address: \tt kolitch@nsfitp.itp.ucsb.edu}\\}
\address{Department of Physics\\University of California\\
		Santa Barbara, CA 93106-9530}

\date{February~~~, 1995}

\maketitle
\begin{abstract}
Solutions to flat space Friedmann-Robertson-Walker cosmologies in
Brans-Dicke theory with a cosmological constant are investigated.
The matter is modelled as a $\gamma$-law perfect fluid.  The field
equations are reduced from fourth order to second order through a change
of variables, and the resulting two-dimensional system is analyzed using
dynamical system theory.  When the Brans-Dicke coupling constant is
positive $(\omega > 0)$, all initially expanding models approach
exponential expansion at late times, regardless of the type of matter
present.  If $\omega < 0$, then a wide variety of qualitatively distinct
models are present, including nonsingular ``bounce'' universes,
``vacillating'' universes and, in the special case of $\omega = -1$,
models which approach stable Minkowski spacetime with an exponentially
increasing scalar field at late times.  Since power-law solutions do not
exist, none of the models appear to offer any advantage over the standard
deSitter solution of general relativity in achieving a graceful exit from
inflation.

\end{abstract}
\vfill\eject

\def\fig#1{Fig.~{#1}}			
\def\Equation#1{Equation~\(#1)}		
\def\Equations#1{Equations~\(#1)}	
\def\Eq#1{Eq.~\(#1)}			
\def\Eqs#1{Eqs.~\(#1)}			
\let\eq=\Eq\let\eqs=\Eqs		
\def\(#1){(\call{#1})}
\def\call#1{{#1}}
\def\square{\kern1pt\vbox{\hrule height 1.2pt\hbox{\vrule width 1.2pt\hskip 3pt
   \vbox{\vskip 6pt}\hskip 3pt\vrule width 0.6pt}\hrule height 0.6pt}\kern1pt}

\section{Introduction}
\label{intro}

A recent renewal of interest in Brans-Dicke (BD) theory\cite{BD} can be traced
to the discovery by La and Steinhardt that the use of BD theory in place
of general relativity can ameliorate the exit problem of inflationary
cosmology\cite{La}.  This is possible because the interaction of the BD
scalar  field with the metric slows the expansion from exponential to
power-law.  Although the original ``extended inflation'' scenario appears
to have been ruled out by observational constraints\cite{LL}, models
which survive include those based on more general scalar-tensor theories,
such as ``hyperextended inflation''\cite{S-A}, and hybrid models which include
both a first-order phase transition and a period of slow-roll\cite{Holman2}.
The renewed interest in scalar-tensor gravitation has led to several recent
investigations into the generation of exact solutions for cosmology in such
theories\cite{Barrow93}, as well as to some qualitative studies of the
models which result\cite{Barrow90,Burd,Damour93,Romero,Wands,Kolitch94}.
It has also been pointed out recently that an inflationary era may result
directly from the dynamics of the scalar field, without any potential or
cosmological constant being necessary\cite{Levin}.

In this paper we are concerned with the behavior of homogeneous and
isotropic cosmological models in Brans-Dicke gravity, with the addition
of a positive cosmological constant.  This differs from standard
extended inflationary scenarios in that the vacuum energy is decoupled
from the scalar field.  The goal is simply to analyze the cosmological
models which such a theory gives rise to, with an emphasis on the question
of whether a viable inflationary model might exist.  Previously, similar
analyses have been performed for the case $\Lambda \ne 0$ with no matter
present\cite{Romero}, and for the case $\Lambda = 0$ with additional
matter present\cite{Wands,Kolitch94}.  The treatment will closely
parallel that given in Ref.~\cite{Kolitch94}, and the interested reader
is referred to that paper for more detail.

In Sec. II, it is
shown that the field equations for this theory can be reduced to a
two-dimensional dynamical system in the case of flat space.  In Sec. III,
the equilibrium points are found and the corresponding solutions
are discussed.  In Sec. IV we summarize the results.

\section{The Field Equations}
\label{BDcosmo}

In this section the field equations are reduced to a planar dynamical
system through a change of variables.  For notation and conventions,
the reader is referred to the parallel treatment given in\cite{Kolitch94}.
Generalizing the action for Brans-Dicke theory to include a nonzero
cosmological constant, it may be written as
\begin{equation}
	S_{BD} = \int d^4x \sqrt{-g} \left(-\phi [R - 2\Lambda]
		+ \omega {{\phi^{,\mu}
		\phi_{,\mu}}\over {\phi}} + 16\pi
		{\cal \char'114}_m \right).
\label{BDaction}
\end{equation}
Taking $\Lambda$ and $\omega$ constant, and varying this action with
respect to the metric and the scalar field, one finds that the nontrivial
components of the field equations in a homogeneous and isotropic (FRW)
spacetime are
\begin{eqnarray}
	\left({\dot a \over a} + {\dot \phi \over {2\phi}} \right)^2
		+ {k \over a^2} =&& \left({2\omega + 3} \over 12 \right)
		\left({\dot \phi \over \phi} \right)^2 +
		{8\pi \rho \over {3\phi}} + {\Lambda \over 3},
		\label{BDFRW1}						\\
	-{1 \over {a^3}}{d \over dt}(\dot \phi a^3) =&&
		\left({8\pi} \over {3 + 2\omega} \right)
		\left(T^\mu{}_\mu - {{\Lambda \phi}\over{4\pi}}\right),
		\label{BDFRW2}
\end{eqnarray}
where $a(t)$ is the cosmic scale factor.
Assuming a perfect fluid form for the stress-energy tensor,
{\it i.e.,\ } $T_{\mu \nu} = \hbox{diag}(\rho, p, p, p)$, the usual
conservation equation is also satisfied (the zeroth component of
$T^{\mu \nu}{}_{;\nu} = 0$):
\begin{equation}
	\dot \rho = -3 {\dot a \over a} (p + \rho).
\label{BDFRW3}
\end{equation}
Assuming only that $\rho > 0$ and $\phi > 0$, inspection of
Eq.~\(\ref{BDFRW1}) reveals that we must have ${\omega\ge-3/2}$ in order
to satisfy the field equations for all values of $k$ and $\Lambda$.
Furthermore, we see from the form of the action in Eq. (\ref{BDaction})
that the integrity of the theory is lost if $\omega = 0$.  We therefore
take ${\omega\ge -3/2}$ and $\omega \ne 0$ in what follows.

Now we take $k=0$, and transform the fourth-order system specified
by Eqs.~\(\ref{BDFRW1}-\ref{BDFRW3}) into a pair of coupled second-order
equations in which, however, only first derivatives of the new variables
appear.  It will be sufficient for our purposes to consider only the models
with $k=0$, as any candidate for a viable inflationary model must at the
very least solve the flatness problem.  First, define the new variables
\begin{eqnarray}
	\Theta \equiv&& \left({\dot a \over a} + {\dot \phi \over {2\phi}}
		\right),
		\label{betadef}						\\
	\Sigma \equiv&& A {\dot \phi \over \phi},
		\label{sigdef}
\end{eqnarray}
where dots represent derivatives with respect to time, and $A \equiv
(2\omega+3)/ 12$.  Next, parametrize the equation of
state by writing $p = (\gamma - 1) \rho$, where, for example,
\begin{equation}
	\gamma = \cases{  0,   & false-vacuum energy; \cr
			  1,   & pressureless dust; \cr
			4/3,   & radiation; \cr
			  2,   & ``stiff'' matter. \cr}
\end{equation}
Finally, take $k=0$ and rewrite Eqs.\ (\ref{BDFRW1}-\ref{BDFRW3})
in terms of these new variables.  Then straightforward differentiation
and resubstitution lead to the equivalent field equations
\begin{eqnarray}
	\dot \Sigma =&& \Theta^2(1-3\gamma /4)-{{\Sigma^2}\over 2A}
		(1-3\gamma /2) - 3\Theta \Sigma - {\Lambda \over 6}
		(1-3\gamma /2),
		\label{DS1}						\\
	\dot \Theta =&& 3{{\Sigma^2}\over A}(\gamma /2 - 1) -
		{{3\gamma\Theta^2}\over 2} + {{\Theta\Sigma}\over {2A}}
		+ {{\Lambda\gamma}\over 2}.
		\label{DS2}
\end{eqnarray}
Eqs.~\(\ref{DS1}) and (\ref{DS2}) constitute a planar dynamical system in the
variables $\Theta$ and $\Sigma$, and are the desired results of this section.

\section{The Equilibrium Points}
\label{qualanal}

The equilibrium points of the dynamical system are obtained by setting
$\dot\Theta$ and $\dot\Sigma$ equal to zero in Eqs.~\(\ref{DS1}) and
(\ref{DS2}), and then solving the resulting equations for $\Theta$ and
$\Sigma$.  As each of these equations is a second order polynomial,
we expect in general four equilibrium points.  When \hbox{$\dot\Theta =
\dot\Sigma = 0$,} this implies that $\dot a / a = \Theta - \Sigma/2A$
and $\dot\phi / \phi = \Sigma/A$ are both constants, so that
\begin{eqnarray}
	a(t) =&& a_0 \exp\left[\left(\Theta - {\Sigma \over {2A}}\right)t
		\right],					   	\\
	\phi(t) = && \phi_0 \exp\left({\Sigma t}\over A \right).
\label{EQa}
\end{eqnarray}
Therefore a fixed point in the $\Theta$--$\Sigma$ system
represents deSitter spacetime, with the addition of an exponentially
varying scalar field.  In the special case $\Theta = \Sigma /2A$, the
solution is not deSitter but rather Minkowski spacetime with a scalar
field.  Note also that only equilibrium points at finite values have been
considered. The global picture, including the behavior of $\Theta$ and
$\Sigma$ at infinity, may be obtained by various compactification
methods\cite{DSref}.  Such an analysis is not necessary for our purposes,
however, as we are primarily interested in whether viable inflationary
models exist in this theory.  In particular, one sees by inspection of
Eqs.~\(\ref{DS1}) and (\ref{DS2}) that the origin of the $\Theta$--$\Sigma$
plane can never be an equilibrium point when $\Lambda\ne 0$.  This
immediately rules out the possiblity of a stable power-law solution,
since such a solution would appear in that plane as line of constant
slope, and would asymptotically approach equilibrium at the origin.

Although solution curves span the entire $\Theta$--$\Sigma$ plane,
the requirement $\rho > 0$, where $\rho$ is the energy density of the
perfect fluid matter, eliminates some regions on physical grounds.
It follows from Eq.~\(\ref{BDFRW1}) with $k=0$ that (assuming $\phi > 0$)
any point $(\Theta_0,\Sigma_0)$ satisfying
	${\Theta_0}^2  > {{\Sigma_0^2} / A} + {\Lambda / 3}$
will lie in $\rho > 0$, whereas points satisfying
$	{\Theta_0}^2  < {{\Sigma_0^2} / A} + {\Lambda / 3}$
lie in $\rho < 0$ and thus do not represent physical solutions.
Now let us proceed with the analysis.  Although we restrict ourselves to
the consideration of models with $\Lambda > 0$, it is clear that
the techniques can easily be extended to models with a negative
cosmological constant.

One pair of equilibrium points are always present regardless
of the value of $\gamma$; they satisfy the field equations for
$\rho = 0$ and are thus vacuum solutions.  These points are
\begin{equation}
	(\Theta_0,\Sigma_0)_{1,2} = \pm\left[{\Lambda (2\omega +3)}\over
		{2(3\omega +4)}\right]^{1/2}(1,1/6)
\end{equation}
and they represent the solutions
\begin{eqnarray}
	a(t) =&& a_0 \exp\left\{\pm(\omega +1)\left[{{2\Lambda}\over
		{(2\omega +3)(3\omega +4)}}\right]^{1/2}t\right\},
		\label{vac1}						\\
	\phi(t) =&& \phi_0\exp\left\{\pm\left[{{2\Lambda}\over
		{(2\omega +3)(3\omega +4)}}\right]^{1/2}t\right\},
		\label{vac2}
\end{eqnarray}
These solutions have previously been noted in the
literature\cite{Barrow90,Romero}, and are attractors for most, but not all,
of the initially expanding models in this theory ({\it cf.} discussion below).
Note that if $\omega = -1$, then there are solutions where the geometry
is Minkowski and the scalar field either grows or shrinks exponentially.
Under the field redefinition $\phi \to e^{-\Phi}$, these can be identified
as the static ``linear dilaton'' solutions of string cosmology\cite{Myers}.

The location and stability of the remaining equilibrium points depends
upon the value of $\gamma$, as well as upon $\omega$ and $\Lambda$.
Hence it is convenient to classify the models, and to discuss the overall
character and stability of the solutions, according to the equation of
state of the perfect fluid.  The additional equilibrium points are as
follows:

\begin{eqnarray}
	(\gamma = 0): \quad (\Theta_0,\Sigma_0)_{3,4} =&&
		\pm\left[\Lambda\over 6\right]^{1/2}(1,0)		\\
	(\gamma = 1): \quad (\Theta_0,\Sigma_0)_{3,4} =&&
		\pm\left[\Lambda\over {6(3\omega +4)}\right]^{1/2}
		\left(1,{{2\omega +3}\over 2}\right)			\\
	(\gamma = 4/3): \quad (\Theta_0,\Sigma_0)_{3,4} =&&
		\pm\left[\Lambda\over {2(2\omega +3)}\right]^{1/2}
		\left(1,{{2\omega +3}\over 3}\right)			\\
	(\gamma = 2): \quad (\Theta_0,\Sigma_0)_{3,4} =&&
		\pm\left[{2\Lambda}\over 3\right]^{1/2}
		\left(1,{{2\omega +3}\over 4}\right)
\end{eqnarray}

The details of the stability analysis are
given in Appendix A.  Table 1 of that appendix lists the eigenvalues
of the equilibrium points, and Table 2
explicity states the existence and stability of each equilibrium point
as a function of $\omega$.  Note that in some cases, the solutions
represented by a given equilibrium point require negative energy
density, and are therefore physically uninteresting.  In such cases we
have simply written ``$\rho < 0$'' in the tables.

The overall character of the solutions may be further examined by
numerically integrating the solutions $\Theta(t)$ and $\Sigma(t)$ for
a variety of initial conditions with each qualitatively distinct set
of parameters $(\Lambda,\omega,\gamma)$.  Figures 1--4 show the results of
this procedure, where we have selected $\Lambda = 3$ arbitrarily.
The shaded regions in each case require $\rho < 0$, and so are disallowed
physically.  Curves to the right of the line $da/dt = 0$ represent expanding
universes, and those to its left represent contracting universes.
Note that if $-1 < \omega < 0$, as for example in Fig. 1b, then there
exist nonsingular ``bounce'' models which pass smoothly from contraction
to expansion, and ``vacillating'' models which pass from expansion to
contraction to reexpansion, or the time reversal of this behavior.
There also exist extreme cases of ``vacillation''; some models can
vacillate several times before settling down to exponential contraction
at late times, as shown in Fig. 4d.

\section{Summary}
\label{concl}

We have shown that in Brans-Dicke theory with a positive cosmological
constant, a wide range of flat-space models exist, including some with
no analogues in general relativity.  In general, these models are
parametrized by the initial conditions of the scalar field, the value
of the BD coupling constant, an initial expansion rate, an equation of
state for the matter and the value of the cosmological constant.  The
first two parameters are, of course, absent in general relativity.
Most, but not all, of the expanding models asymptotically approach
vacuum deSitter spacetime at late times.  Power-law expansion is not
possible when $\Lambda$ is nonzero.

If $\omega > -1$, there exist two finite-valued equilibrium points of
physical interest, representing the vacuum solutions (\ref{vac1})
and (\ref{vac2}).  One of these is stable and corresponds to expanding
deSitter spacetime, and all initially expanding models approach this solution
asymptotically.  As $\omega \to \infty$, the solution becomes identical
to the deSitter universe where $a(t) \sim e^{\Lambda t / 3}$ and
$\phi(t) = \hbox{constant}$, in accordance with the well-established
correspondence between GR and BD theory in this limit.  When $\omega > 0$,
all contracting models contract to a singularity; however, if
$-1 < \omega < 0$, then nonsingular ``bounce'' models are also possible,
and these may or may not approach deSitter spacetime at late times,
depending upon the initial conditions of the model.  This behavior has been
noted by other authors\cite{bounce}.  Also there are ``vacillating'' models,
which expand from a big bang, slow down, and recontract before continuing
their expansion and approaching deSitter spacetime.  The time-reversal of
this behavior also exists.

If $\omega = -1$, there exist ``static-exponential'' solutions, where the
geometry is Minkowski spacetime while the scalar field changes exponentially
with time.  The solution with increasing scalar is found to be stable, and
all models which expand from a big bang approach it at late times, regardless
of the type of matter present.  The stability of these models may be
explained by the fact that in the Newtonian limit,
$\phi \sim G^{-1}$\cite{SW}.  Hence the exponential increase in the
scalar field corresponds to an exponential weakening of the gravitational
interactions, ensuring that the universe does not recollapse regardless
of its matter content.

If $-3/2 < \omega < -1$, then the behavior of the models depends upon the type
of matter present, and we can distinguish two classes of behavior.  {\bf (i)}
In the cases of false-vacuum energy and pressureless dust, only the equilibrium
points representing the vacuum solutions (\ref{vac1}) and (\ref{vac2}) are
present in the regime $\rho > 0$, and the contracting solution
is stable.  All models which start from a big bang eventually contract to
a singularity; this collapse will become asymptotically exponential
if $-4/3 < \omega < -1$, or superexponential if $-3/2 < \omega < -4/3$.
There also exist models which start with a finite rate of expansion
and expand perpetually.  {\bf (ii)}  In the cases of radiation and ``stiff''
matter, the dynamical system undergoes a qualitative change at a particular
value of $\omega$.  This critical value is $\omega_c = -5/4$ for radiation,
and $\omega_c = -7/6$ for ``stiff'' matter.  If $\omega_c \le \omega < -1$,
then only the equilibrium points representing (\ref{vac1}) and (\ref{vac2})
are present with $\rho \ge 0$, and the contracting solution is stable.
If $-4/3 < \omega < \omega_c$, then there are four equilibrium
points in the regime $\rho \ge 0$, representing both vacuum and
non-vacuum deSitter solutions; however, only the contracting non-vacuum
solution is stable.  In these solutions, the exponential growth of the
scalar field exactly balances that of the energy density, so that the
ratio $\rho / \phi$ is constant.  Thus the ordinary matter acts exactly
like a cosmological constant, since it is this ratio which appears as the
matter source term in the field equation (\ref{BDFRW1}).  All models
will collapse to a singularity; the collapse will be asymptotically
exponential for models starting from a big bang.  If $-3/2 < \omega < -4/3$,
the unstable vacuum equilibrium points no longer exist; otherwise the
behavior of the models remains unchanged.

Although somewhat exotic, the models discussed here do not seem to have any
hope of solving the graceful exit problem of inflationary cosmology.
In models of extended inflation, the mediation of the scalar field slows
the expansion from exponential to power-law, so that the Hubble parameter
decreases with time and true-vacuum bubble nucleation may complete the
inflation-ending phase transition.  Here, however, the cosmological
constant induces deSitter spacetime, and all of the problems of Guth's
``old'' inflation recur.  In addition, one is faced with the question of
the origin of the cosmological constant in these models.  Although there
is no {\it a priori} reason to exclude such a term from the field equations,
the usual explanation of a field with a nonzero potential as the source
of the vacuum energy is not available in this case, since such a potential
term would be coupled to the scalar field.

\section{Acknowledgements}
This research was supported in part by the National Science Foundation under
Grants No.~PHY89-04035 and PHY90-08502.

\appendix
\section{}

Defining $\xi^{(1)} \equiv \Theta - \Theta_0 \quad {\rm and} \quad \xi^{(2)}
\equiv \Sigma - \Sigma_0$, the dynamical system specified by (\ref{DS1}) and
(\ref{DS2}) may be written in the form
\begin{equation}
	{{d \roarrow{\xi}} \over {dt}} = {\bf {\rm J}} \roarrow{\xi}
		\quad + \quad \dots \quad ,
\end{equation}
where the Jacobian is
\begin{equation}
	{\bf {\rm J}} =
	\pmatrix{-3\gamma\Theta_0 + \Sigma_0 / 2A
		&-6(\gamma /2 - 1)\Sigma_0 / A
		+ \Theta_0 / A					\cr
		\noalign{\medskip}
		2\Theta_0(1-3\gamma /4) - 3\Sigma_0
		&-(1-3\gamma /2)\Sigma_0 / A - 3\Theta_0 	\cr}.
\end{equation}
\smallskip\noindent

In cases where the eigenvalues of the Jacobian all have
nonvanishing real part, the fixed point is called hyperbolic and we can
determine its stability from the signs of those real parts: if the real
part of each of the eigenvalues is negative at a given equilibrium point,
the solution is stable at that point; if the real part of each eigenvalue
is positive, or if the real part of one eigenvalue is positive and that
of the other is negative, then the solution is unstable at that point.
Finally, if the real part of any of the eigenvalues is zero at a point,
then the point is called nonhyperbolic and its stability in the
neighborhood of that point cannot be determined by this method
\cite{DSref}.

Table 1 shows the eigenvalues of the Jacobian for each equilibrium point,
and Table 2 explicitly states the existence and stability of the
equilibrium points as a function of $\omega$.  In cases where the
solution represented by the equilibrium point requires negative energy
density, we have simply written ``$\rho < 0$''.  The points are labelled
in accordance with the conventions in the text, {\it i.e.,\ } for each
value of $\gamma$, points 1 and 2 represent vacuum solutions, and points
3 and 4 represent non-vacuum solutions.
\begin{table}
\caption{Eigenvalues of the Jacobian Matrix for BD Cosmology with
$\Lambda > 0$}
\begin{tabular}{ccc}
Equilibrium Point & $\lambda_1$ & $\lambda_2$\\
\tableline
$(\gamma = 0)_1$ & $-[2\Lambda / (2\omega +3)(3\omega + 4)]^{1/2}$
	& $-[2\Lambda(3\omega +4) / (2\omega +3)]^{1/2}$\\
$(\gamma = 0)_2$ & $+[2\Lambda (3\omega +4) / (2\omega +3)]^{1/2}$
	& $+[2\Lambda / (2\omega +3)(3\omega + 4)]^{1/2}$\\
$(\gamma = 0)_3$ & $\rho < 0$ & $\rho < 0$\\
$(\gamma = 0)_4$ & $\rho < 0$ & $\rho < 0$\\
$(\gamma = 1)_1$ & $-[2\Lambda(3\omega +4) / (2\omega +3)]^{1/2}$
	& $-[2\Lambda(3\omega +4) / (2\omega +3)]^{1/2}$\\
$(\gamma = 1)_2$ & $+[2\Lambda(3\omega +4) / (2\omega +3)]^{1/2}$
	& $+[2\Lambda(3\omega +4) / (2\omega +3)]^{1/2}$\\
$(\gamma = 1)_3$ & $\rho < 0$ & $\rho < 0$\\
$(\gamma = 1)_4$ & $\rho < 0$ & $\rho < 0$\\
$(\gamma = 4/3)_1$ & $-[2\Lambda(3\omega +4) / (2\omega +3)]^{1/2}$
	& $-(4\omega +5)[2\Lambda / (2\omega +3)(3\omega +4)]^{1/2}$\\
$(\gamma = 4/3)_2$ & $+(4\omega +5)[2\Lambda / (2\omega +3)(3\omega +4)]
	^{1/2}$ & $+[2\Lambda(3\omega +4) / (2\omega +3)]^{1/2}$\\
$(\gamma = 4/3)_3$
	& $C(-1 + \sqrt{64\omega + 81})$\tablenotemark[1]
	& $C(-1 - \sqrt{64\omega + 81})$\tablenotemark[1]\\
$(\gamma = 4/3)_4$
	& $C(+1 + \sqrt{64\omega + 81})$\tablenotemark[1]
	& $C(+1 - \sqrt{64\omega + 81})$\tablenotemark[1] \\
$(\gamma = 2)_1$ & $-[2\Lambda(3\omega +4) / (2\omega +3)]^{1/2}$
	& $-(6\omega +7)[2\Lambda / (2\omega +3)(3\omega +4)]^{1/2}$\\
$(\gamma = 2)_2$ & $+(6\omega +7)[2\Lambda / (2\omega +3)(3\omega +4)]
	^{1/2}$ & $+[2\Lambda(3\omega +4) / (2\omega +3)]^{1/2}$\\
$(\gamma = 2)_3$
	& $C(-\sqrt{3(2\omega +3)} + \sqrt{102\omega + 121})
		$\tablenotemark[1]
	& $C(-\sqrt{3(2\omega +3)} - \sqrt{102\omega + 121})
		$\tablenotemark[1]\\
$(\gamma = 2)_4$
	& $C(+\sqrt{3(2\omega +3)} + \sqrt{102\omega + 121})
		$\tablenotemark[1]
	& $C(+\sqrt{3(2\omega +3)} - \sqrt{102\omega + 121})
		$\tablenotemark[1]\\
\end{tabular}
\label{table1}
\tablenotetext[1]{$C \equiv [\Lambda / 2(2\omega +3)]^{1/2}$}
\end{table}

\begin{table}
\caption{Existence and Stability of the Equilibrium Points}
\begin{tabular}{cccc}
Equilibrium Point & $-3/2 < \omega \le -4/3$ & $-4/3 < \omega
< \omega_c$\tablenotemark[1] & $\omega \ge \omega_c$\\
\tableline
$(\gamma = 0,\,1)_1$ & nonexistent & N/A & stable\\
$(\gamma = 0,\,1)_2$ & nonexistent & N/A & unstable\\
$(\gamma = 0,\,1)_3$ & $\rho < 0$ & N/A & $\rho < 0$\\
$(\gamma = 0,\,1)_4$ & $\rho < 0$ & N/A & $\rho < 0$\\
$(\gamma = 4/3,\,2)_1$ & nonexistent & unstable & stable\\
$(\gamma = 4/3,\,2)_2$ & nonexistent & unstable & unstable\\
$(\gamma = 4/3,\,2)_3$ & stable & stable & $\rho < 0$\\
$(\gamma = 4/3,\,2)_4$ & unstable & unstable & $\rho < 0$\\
\end{tabular}
\label{table3}
\tablenotetext[1]{$\omega_c(\gamma = 0,1) = -4/3$\, ; $\omega_c(\gamma = 4/3)
= -5/4$\, ; $\omega_c(\gamma = 2) = -7/6$}
\end{table}

\gdef\journal#1, #2, #3, 1#4#5#6{		
    {\sl #1~}{\bf #2}, #3 (1#4#5#6)}		

\def\pr{\journal Phys. Rev., }

\def\pra{\journal Phys. Rev. A, }

\def\prb{\journal Phys. Rev. B, }

\def\prc{\journal Phys. Rev. C, }

\def\prd{\journal Phys. Rev. D, }

\def\prl{\journal Phys. Rev. Lett., }

\def\jmp{\journal J. Math. Phys., }

\def\rmp{\journal Rev. Mod. Phys., }

\def\cmp{\journal Comm. Math. Phys., }

\def\np{\journal Nucl. Phys., }

\def\pl{\journal Phys. Lett., }

\def\apj{\journal Astrophys. Jour., }

\def\apjl{\journal Astrophys. Jour. Lett., }

\def\annp{\journal Ann. Phys. (N.Y.), }

\begin{figure}
\caption{The evolution of solutions in the $\Theta$--$\Sigma$ plane with
	$\gamma = 0$ (false-vacuum energy), where $\Lambda = 3$ has been
	selected for convenience.
	(a) $\omega = 9/2$, representative of $\omega > 0$.
	Solutions to the right of the line $da/dt = 0$
	represent expanding universes; those to the left are contracting
	universes.  The shaded region requires negative energy density
	and so is disallowed physically.  (b) $\omega = -1/2$,
	representative of $-1 < \omega < 0$.  The line $da/dt = 0$
	has moved into the regime of positive energy density, so that
	some models pass smoothly from contraction to expansion, a
	general feature of models with $\omega < 0$.  (c) $\omega = -1$.
	Note the attractive nature of the ``static-exponential'' solution.
	(d) $\omega = -1.4$, representative of $-3/2 < \omega < -4/3$.
	All models collapse to a singularity regardless of initial
	conditions.
\label{fig1}}
\end{figure}

\begin{figure}
\caption{Some models with $\gamma = 1$ (pressureless dust) and $\Lambda = 3$.
	(a) $\omega = 9/2$, representative of $\omega > 0$.
	(b) $\omega = -1/2$, representative of $-1 < \omega < 0$.
	(c) $\omega = -1$.
	(d) $\omega = -1.4$, representative of $-3/2 < \omega < -4/3$.
\label{fig2}}
\end{figure}

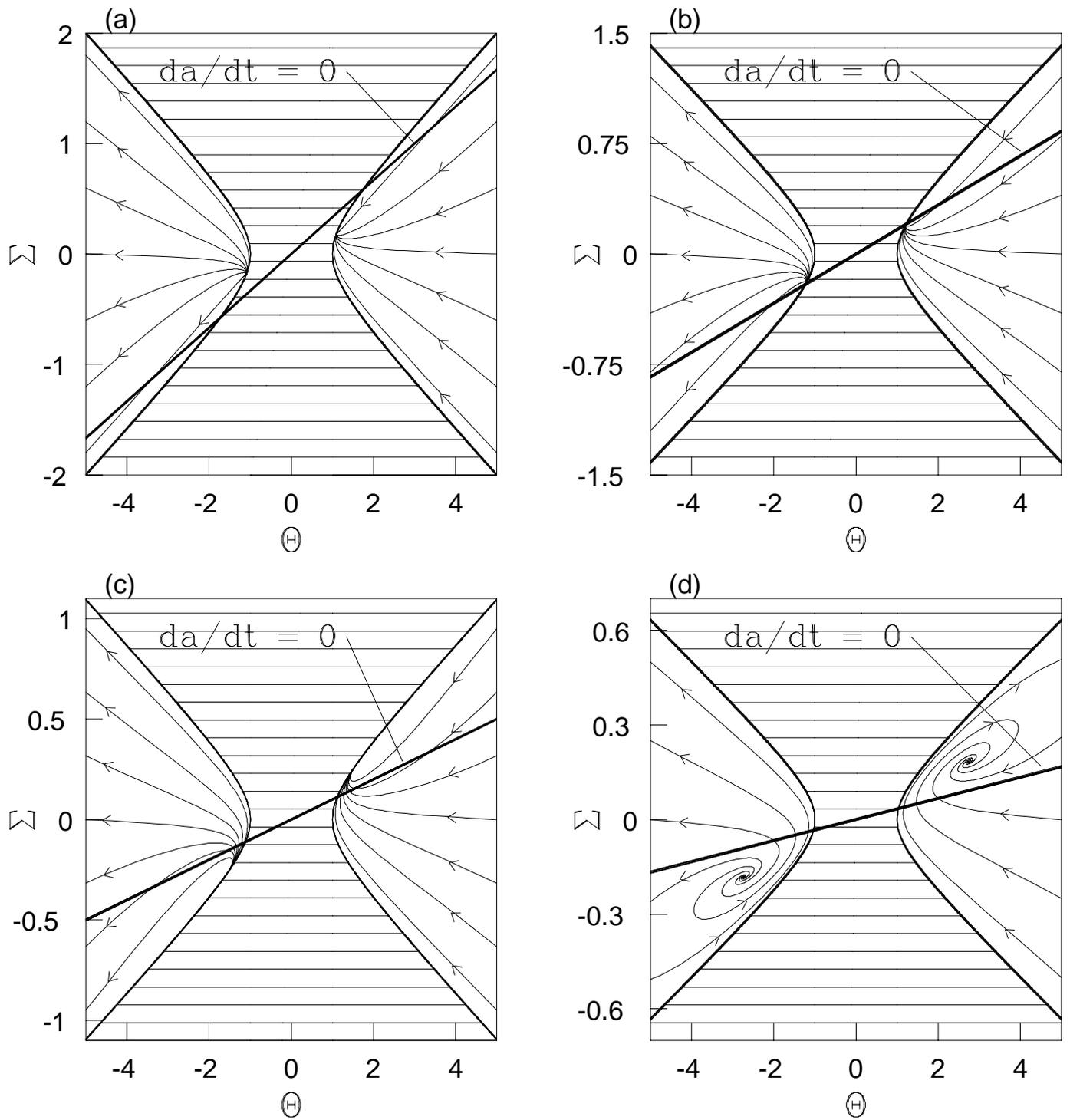
\begin{figure}
\caption{Some models with $\gamma = 4/3$ (radiation) and $\Lambda = 3$.
	(a) $\omega = -1/2$, representative of $-1 < \omega < 0$.
	(b) $\omega = -1$.
	(c) $\omega = -1.2$, representative of $-5/4 \le \omega < -1$.
	(d) $\omega = -1.4$, representative of $-3/2 < \omega < -5/4$.
\label{fig3}}
\end{figure}

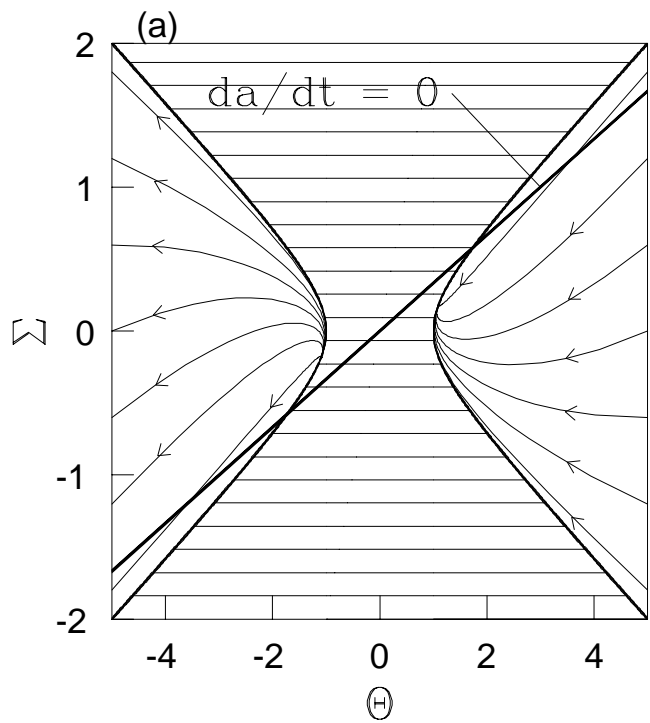
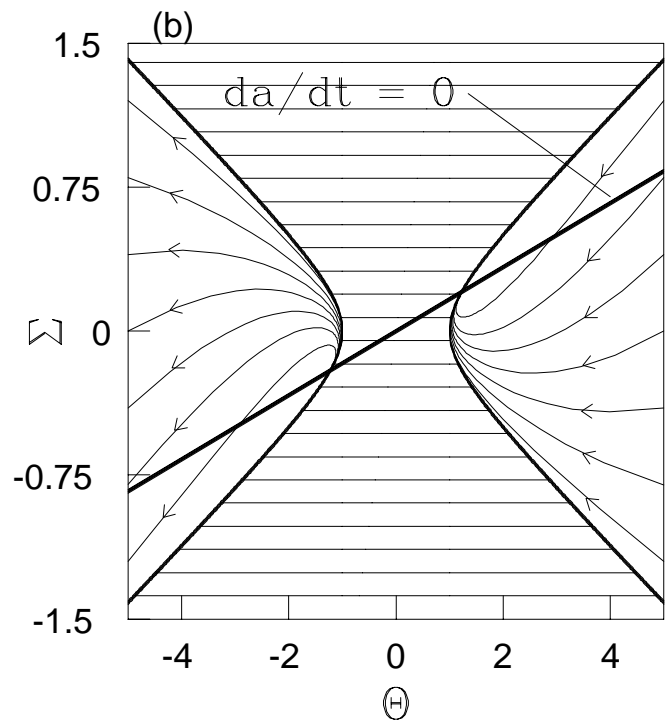
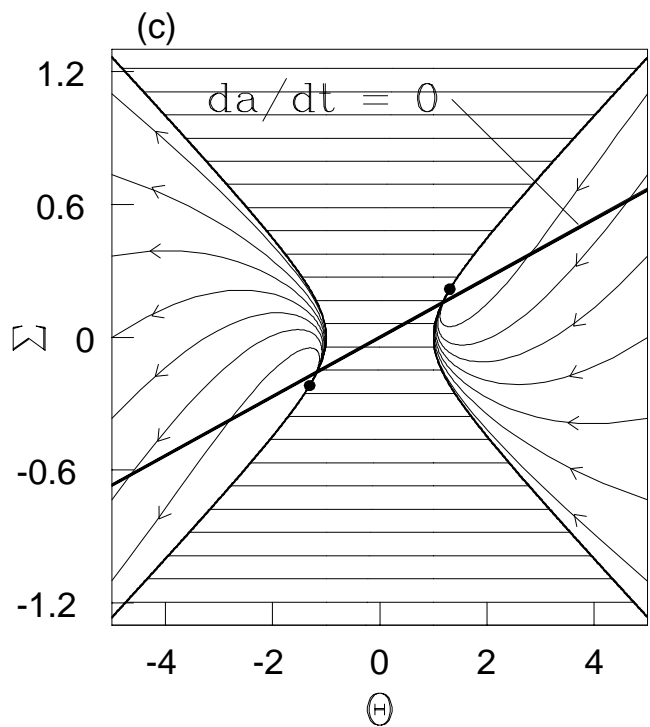
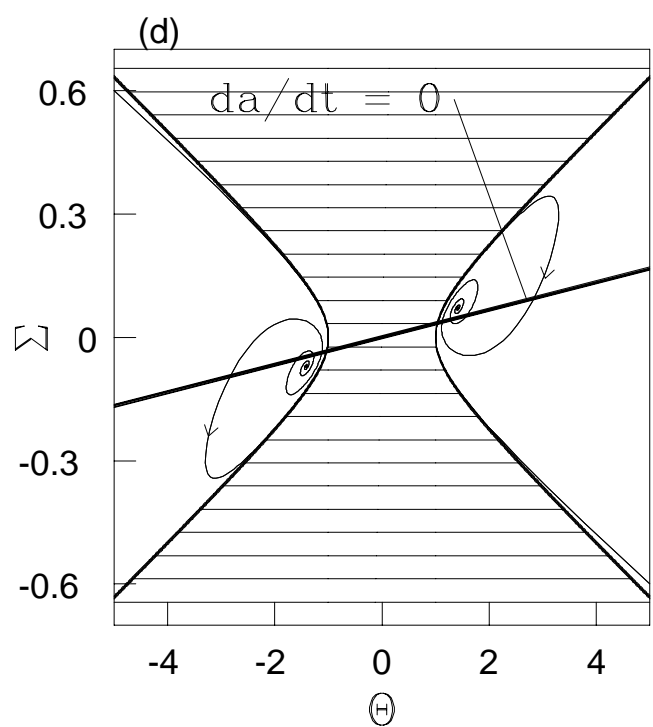
\begin{figure}
\caption{Some models with $\gamma = 2$ (``stiff matter'') and $\Lambda = 3$.
	(a) $\omega = -1/2$, representative of $-1 < \omega < 0$.
	(b) $\omega = -1$.
	(c) $\omega = -1.2$, representative of $-7/6 \le \omega < -1$.
	The locations of the equilibrium points are marked with dots for
	clarity.
	(d) $\omega = -1.4$, representative of $-3/2 < \omega < -7/6$.
	Only one solution is shown to simplify the plot.
\label{fig4}}
\end{figure}

\end{document}